\documentclass{cjpsuppl}
\usepackage{graphicx}

\begin{document}
\title{Heavy Ions --- Prospects at the LHC}
\authori{B. M\"uller}
\addressi{Department of Physics, Duke University, Durham, NC 27708, USA}
\authorii{}    \addressii{}
\authoriii{}   \addressiii{}
\authoriv{}    \addressiv{}
\authorv{}     \addressv{}
\authorvi{}    \addressvi{}
\headtitle{Heavy Ions --- Prospects at the LHC}
\headauthor{B. M\"uller}
\lastevenhead{B. M\"uller: Heavy Ions --- Prospects at the LHC}
\pacs{25.75.Nq,25.75.-q,12.38.Mh}
\keywords{relativistic heavy ions, quark-gluon plasma, LHC}
\refnum{}
\daterec{;\\final version}
\suppl{A}  \year{2004} \setcounter{page}{1}
\maketitle

\begin{abstract}
This is a review of the physics prospects for relativistic heavy ion
collisions in the CERN Large Hadron Collider. The motivation for the 
study of superdense matter created in relativistic heavy ion collision
is the prospect of observing a novel state of strongly interacting matter,
the quark-gluon plasma. Experiments at the CERN Super Proton Synchrotron
(SPS) and the Relativistic Heavy Ion Collider (RHIC) at Brookhaven have
yielded important clues of the characteristic signatures of this new
state. The LHC will extend the range of energy densities that can be
explored and facilitate the observation of plentiful hard probes (jets,
heavy quarks) of the properties of the dense matter.
\end{abstract}

\section{The Quark-Gluon Plasma}

Computer simulations of lattice QCD predict that the properties of
strongly interacting matter change drastically at an energy density
of the order of 1 GeV/fm$^3$ or, for baryon symmetric matter, at the
temperature $T_c=165\pm 10$ MeV [1]. According
to the best available calculations, this transition is not a true
(discontinuous) phase transition at zero net baryon density, or 
baryochemical potenial $\mu_B=0$, but at rapid crossover from a
low-temperature phase dominated by hadrons into a high-temperature
phase best characterized as a strongly interacting plasma of quarks
and gluons. The transition is seen as a steep rise in the vicinity
of $T_c$ in the effective number of degrees of freedom $g_{\rm D}$ 
defined as $\epsilon = g_{\rm D}\pi^2 T^4/30$, where $\epsilon$ is 
the energy density and $T$ the temperature (see Fig.~1).

\begin{figure}[htbp]
\centering
\includegraphics[width=0.8\linewidth]{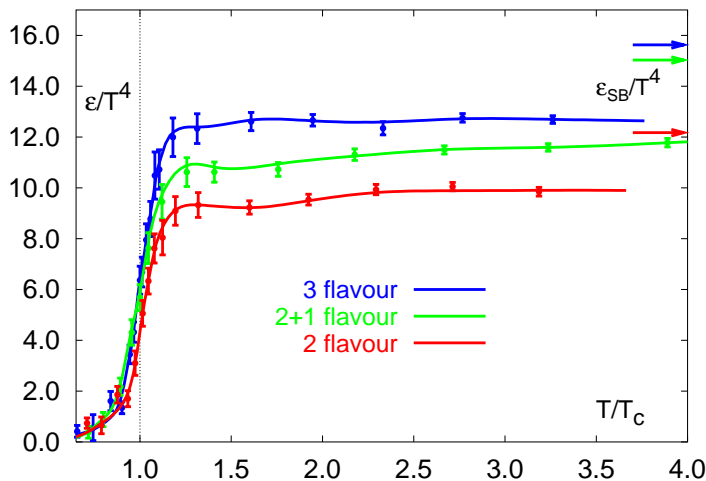}
\caption{Energy density in QCD with 2 and 3 quark flavors, showing 
         a transition temperature $T_c$ between a hadronic gas and 
         a plasma of quarks and gluons (from [2]).}
\end{figure}

For nonzero net baryon density the crossover narrows further and
is expected to become singular at a critical point $(T_{\rm cr},
\mu_{B{\rm cr}})$, beyond which the phase transition is of first order.
Lattice calculations of the location of this critical point have
become possible only within the past two years; the predictions 
are still quite uncertain with a current best estimate of (160, 360)
MeV [3,4]. The change in the properties of the matter at the critical
line $(T_c(\mu_B),\mu_B)$ is characterized by the disappearance of
the vacuum condensate $\langle\bar\psi \psi\rangle$ of light quarks,
accompanied by the screening of the force between colored quanta.

Among the proposed signatures for the formation of a quark-gluon
plasma in relativistic heavy ion collisions are [5]:
\begin{itemize}
\item Effects of the latent heat in the $(\epsilon,T)$ relation;
\item Enhancement of $s$-quark production;
\item Disappearance of light hadrons, especially the $\rho^0$;
\item Bulk hadronization by quark recombination;
\item Disappearance of $(c{\bar c})$ and $(b{\bar b})$ bound states;
\item Large energy loss of fast partons (jet quenching);
\item Thermal $\ell^+\ell^-$ and $\gamma$ radiation;
\item Critical fluctuations (momentum, baryon number);
\item Collective vacuum excitations.
\end{itemize}

Many of the signatures have been observed and studied in heavy ion
collisions at the SPS at CERN and the RHIC at Brookhaven. The
experiments with Pb+Pb collisions at $\sqrt{s_{NN}}=17.6$ GeV at
the SPS showed a significant enhancement of the production of
strange baryons and antibaryons, increasing with their strangeness 
content [6], a suppression in the production of the $J/\Psi$ which 
grows with the centrality of the collision [7], and a strong 
enhancement of low-mass dilepton production below the $\rho^0$ 
resonance, presumably caused by a broadening (or mass shift) of 
the $\rho$-meson in dense matter [8]. In recent years, indications 
of a plateau in the spectral slopes of $K$-mesons over the range
of energies available at the SPS was also found [9].

In its first three years of running, the RHIC program confirmed 
the strong enhancement of strange baryon production and found new,
unexpected evidence for the formation of hadrons out of a bulk
deconfined phase by quark recombination [10]. It also discovered
the strong suppression of high-$p_T$ hadrons in Au+Au collisions
at $\sqrt{s_{NN}}=200$ GeV due to final state interactions [11].
This effect (often called ``jet quenching'') is understood to be
a result of the energy loss of hard scattered partons on their 
passage through the created matter by gluon radiation [12]. For
a detailed evaluation of the results obtained by the experiments
at RHIC, see the collaboration white papers [13].

\section{Heavy Ions in the LHC}

At LHC $^{208}$Pb nuclei will collide with a center-of-mass energy 
$\sqrt{s}=5.5$ TeV. The almost 30-fold increase in center-of-mass energy
over RHIC will lead to a much higher initial energy density and to
an even faster equilibration. Higher energy density and increased 
lifetime of the deconfined phase will enhance the role of the QGP 
phase over final state hadronic interactions. Furthermore, jets and
high-$p_T$ hadrons with transverse momenta of 100 GeV/$c$ and more
will become available [14]. Finally, $c$ and $b$ quarks will become 
plentiful probes at the LHC energy [15].

I will next discuss what we have learned so far from RHIC experiments
about hard QCD probes and give some predictions and extrapolations 
that have been made for the LHC energy. Before doing so, however,
it is important to review the physics of parton saturation, which 
is indispensible when one wants to understand the initial conditions
for nuclear collisions at the LHC.

\subsection{Parton saturation}

The physics of saturation of the nuclear parton distributions [16] 
at small values of the Bjorken variable $x$ has been the subject of
intense research over the past decade [17,18]. Because of its reach
to much lower values of $x$ ($10^{-4} - 10^{-5}$), the LHC will be 
an ideal testing ground for saturation physics. The concept of parton
saturation is based on the notion that the gluon distribution in a 
hadron or nucleus cannot continue to grow faster and faster at small
$x$ without violating unitarity. At some point, gluon fusion must
balance the growth caused by gluon splitting. The scale at which the 
probability of interaction among the partons in the nuclear wave 
function approaches unity determines the saturation scale $Q_s$ [16]:
\begin{equation}
\frac{x G_A(x,Q_s^2)}{\pi R_A^2} \frac{\alpha_s(Q_s^2)}{Q_s^2} \sim 1.
\end{equation}
The universal form of the gluon distribution that emerges when
saturation is reached has been called a {\it color glass condensate}
(CGC). Fits to HERA and NMC data suggest that the saturation scale 
grows with nuclear size and decreasing $x$ as [19]:
\begin{equation}
Q_s^2 = Q_0^2 (A/R_A)^{\delta} (x_0/x)^{\lambda}
\end{equation}
with $\delta=1.266$ and $\lambda=0.288$. Since the charged multiplicity
is related to the saturation scale by $dN_{\rm ch}/dy \sim \pi R_A^2 Q_s^2$,
the saturation picture predicts a three-fold increase in the multiplicity
from RHIC to the LHC, or $dN_{\rm ch}/dy \approx 2000$ in central Pb+Pb
collisions [19,20] (see Fig.~2). The same calculations predict that the 
initially reached energy density of equilibrated gluon matter could be 
as high as 200 GeV/fm$^3$ at a time of 0.2 fm/$c$, corresponding to a 
temperature $T \approx 600$ MeV -- almost twice the value initially 
reached at RHIC.

\begin{figure}[htbp]
\centering
\includegraphics[width=0.8\linewidth]{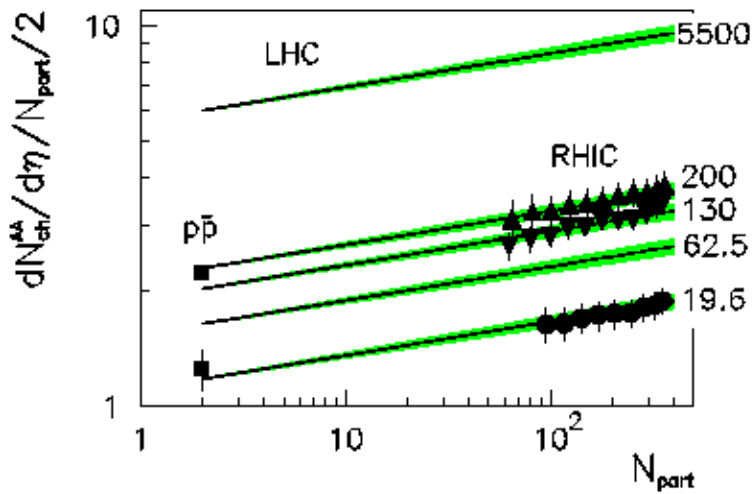}
\caption{Charged particle multiplicity per unit pseudorapidity predicted
         by the gluon saturation model in coparison with SPS and RHIC
         data (from [19]).}
\end{figure}

\subsection{Jet quenching}

Jet quenching is the name for the suppression of hadrons emitted with
large transverse momenta, due to the energy loss of the partonic
progenitors on their passage through the dense quark-gluon plasma.
The name is a bit of a misnomer, because the energy contained in the
jet is not reduced, but only redistributed to smaller values of the
fragmentation variable $z$ and also in the angle with respect to the
jet axis. The theory of this effect is now quite well developed [12]. 
The leading parton radiates gluons induced by soft scattering on the
constituents of the medium. The radiated gluons, in turn, rescatter 
inside the medium, thereby increasing their relative transverse momentum 
$k_T$ with respect to the primary parton. This secondary rescattering 
is peculiar to QCD (rescattering of photons is not important in QED
unless the medium is very thick), and leads to a quadratic dependence
of the energy loss $\Delta E$ on the path length $L$ inside the medium. 
The effect can be expressed as a medium induced modification of the 
fragmentation function $D(z)$ and the angular jet profile [21,22].

It is predicted that jet quenching at the LHC is much larger than 
at the RHIC due to the much higher initial energy density of the 
medium. However, the amount of suppression of the leading hadrons
that can be reached even in the most extreme case is limited, because
partons scattered near the nuclear surface will always escape with
rather little energy loss. High-$p_T$ hadron production then becomes
essentially a surface effect [23].  As the energy density increases, 
the visible surface layer shrinks, and the suppression becomes less
and less sensitive to the high density interior, rendering high-$p_T$
hadrons a somewhat ``fragile'' probe of dense matter [24]. Individual
predictions for the nuclear suppression factor $R_{AA}$ vary somewhat
[24-26] (see Figs.~3 and 4) but, generally, the suppression is not expected 
to be much stronger than at RHIC, where $R_{AA}\approx 0.25$ in central 
Au+Au collisions at the highest measured momenta ($p_T \approx 10$ GeV/c).

\begin{figure}[htbp]
\centering
\includegraphics[width=0.49\linewidth]{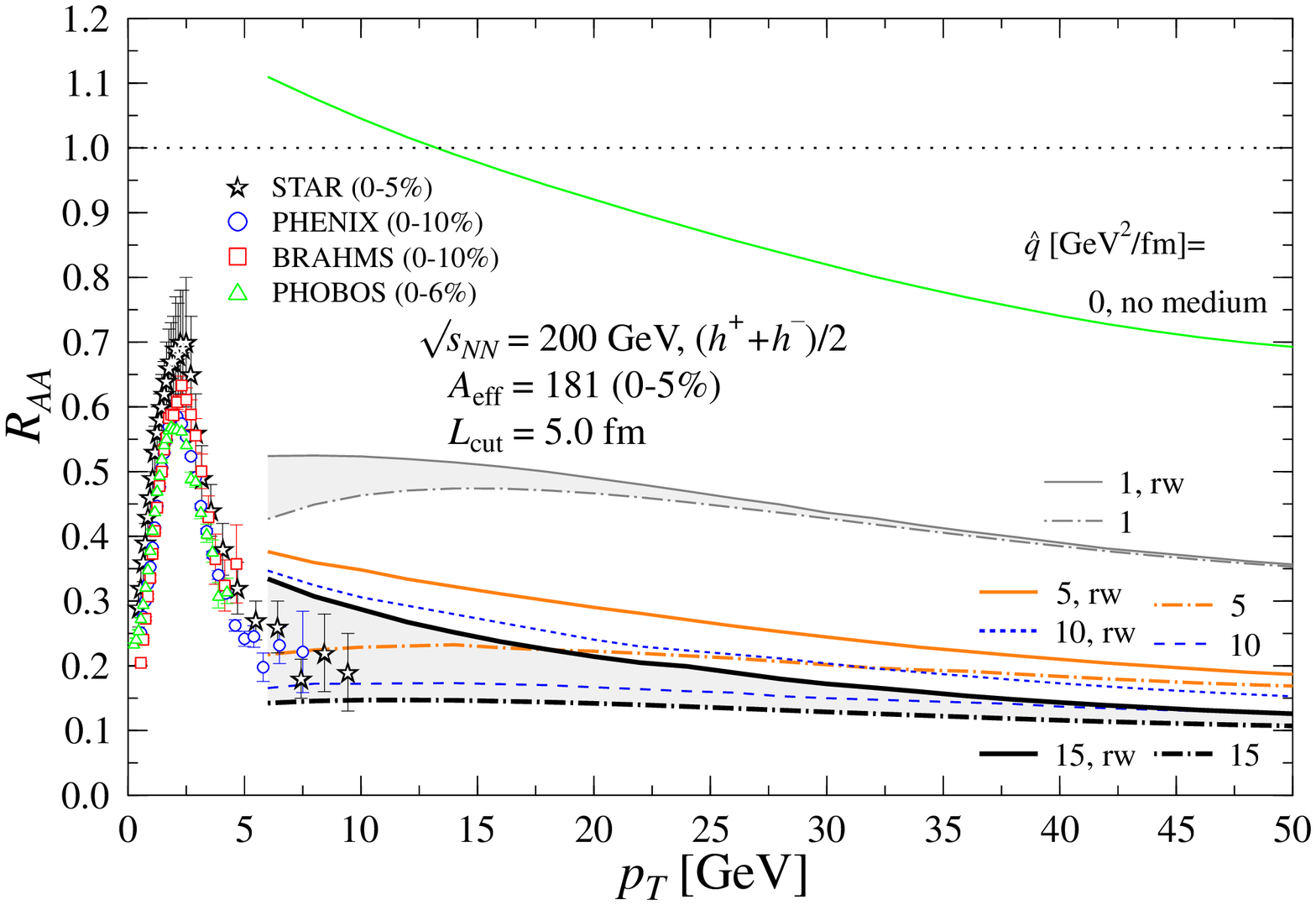}
\includegraphics[width=0.49\linewidth]{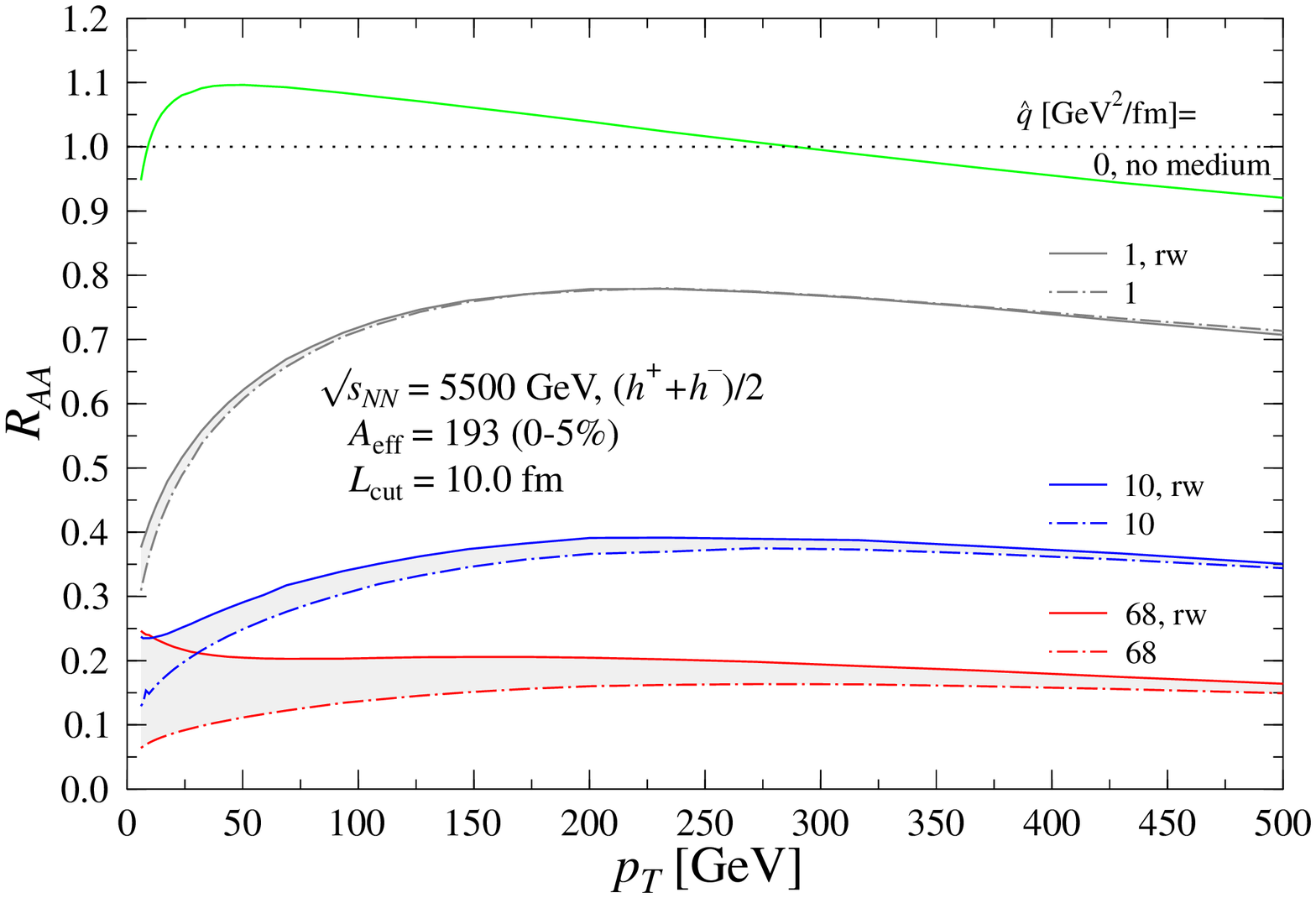}
\caption{Nuclear suppression factor $R_{AA}$ for charged hadron in central
         Au+Au collisions at RHIC (left, with data) and Pb+Pb collisions
         at LHC (right) (from [24]).}
\end{figure}

\begin{figure}[htbp]
\centering
\includegraphics[origin=c,angle=-90,width=0.39\linewidth]{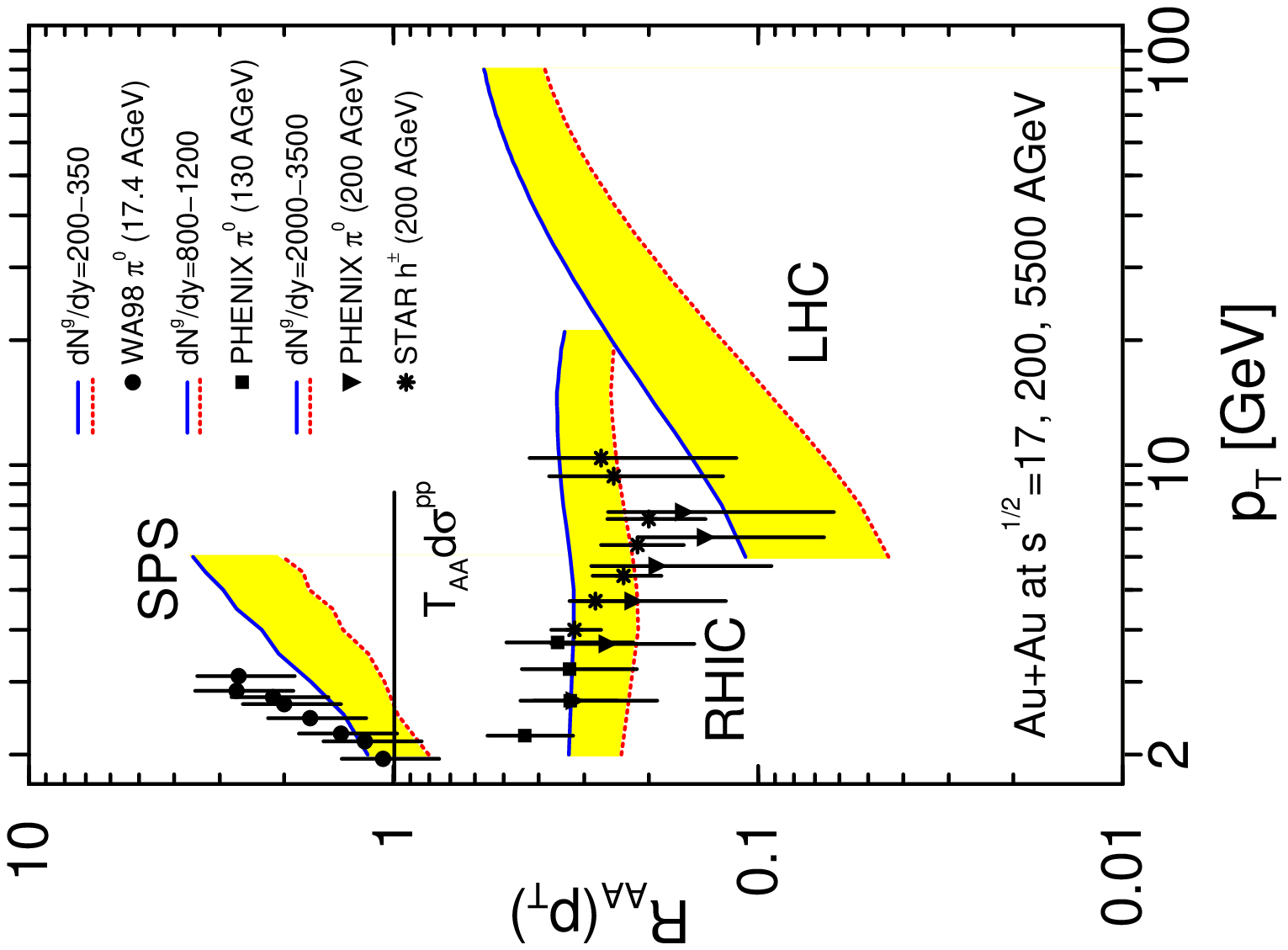}
\includegraphics[width=0.59\linewidth]{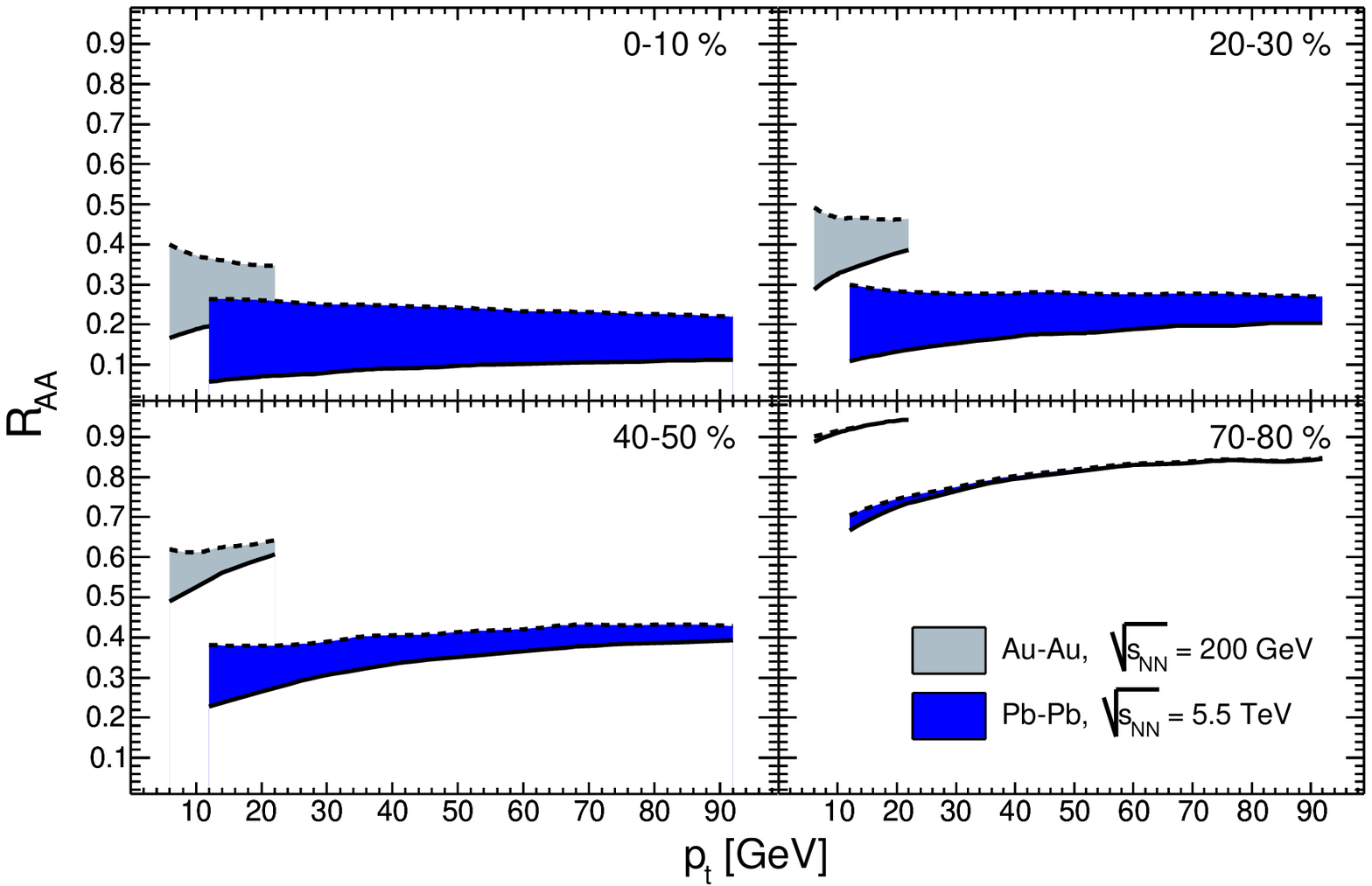}
\caption{Left: Nuclear suppression $R_{AA}$ factor at SPS, RHIC and LHC
         energies for central Au+Au or Pb+Pb collisions (from [25]).
         Right: Centrality dependence of the nuclear suppression factor 
         at RHIC and LHC (from [26]). The suppression disappears faster
         for peripheral collisions at LHC than at RHIC.}
\end{figure}

In contrast to RHIC, it should be possible at the LHC to observe jets 
by calorimetry even in central Pb+Pb collisions. Together with the
hadron spectrum within a jet, this would allow for a direct measurement
of the modification of the fragmentation function and the energy loss 
of the leading parton. The concomitant broadening of the jet itself,
when observed, would constitute a test of the energy loss mechanism by 
gluon radiation. Calculations, shown in Fig.~5, predict that hadrons 
with an intrinsic momentum $k_t < 1$ GeV/$c$ should be suppressed, 
while hadrons with $k_t \sim 4$ GeV/$c$ should be enhanced. Quantitative 
details depend on the jet cone definition [27] .

\begin{figure}[htbp]
\centering
\includegraphics[width=0.7\linewidth]{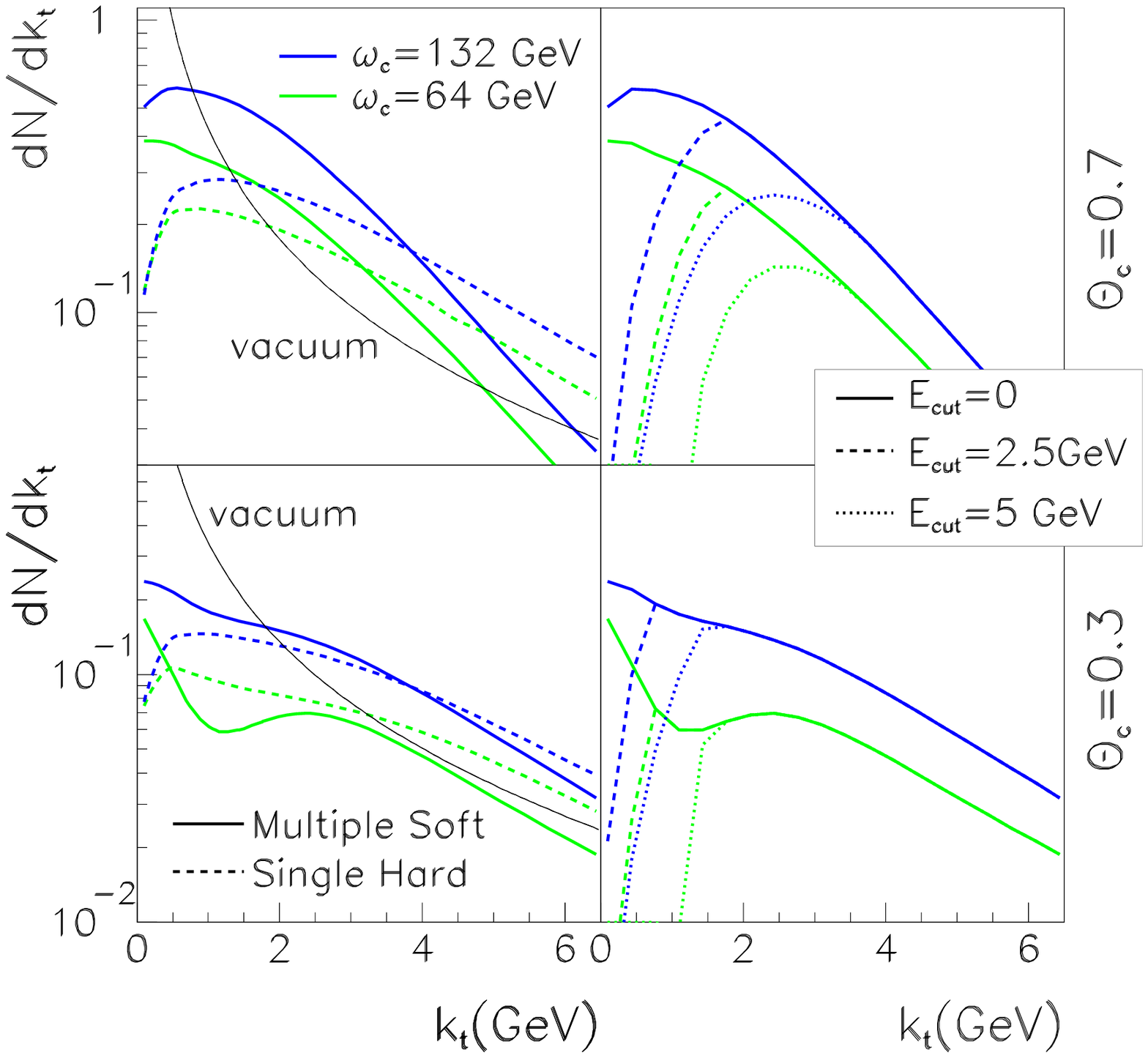}
\caption{Quenching effect on the particle distribution within the jet 
         cone for two different opening angles (from [27]).}
\end{figure}

Other processes, which may become accessible for investigation at
the LHC, are photon tagged jets [28] allowing for a direct measurement
of the energy loss of the leading hadron and parton-to-photon
conversion in the dense medium by Compton backscattering [29].

\subsection{Parton recombination}

One of the surprises found at RHIC was the relative enhancement --
or rather, lack of nuclear suppression -- of baryons in the momentum
range 2 GeV/$c < p_T < 4$ GeV/$c$ [30,31]. This effect has been explained 
as the result of baryon formation by recombination of deconfined quarks
leaving the quark-gluon plasma at the same moment [10,32,33]. It can be 
shown quite generally that recombination dominates over the leading 
twist mechanism of hadron formation, i.\ e.\ fragmentation, when the 
parton spectrum $w_i= e^{-P^+ /T}$ is thermal, independent of the 
details of the hadron wavefunction [32]. The resulting hadrons exhibit
the same thermal distribution, because the exponential factors from
the constituents combine as
\begin{eqnarray}
w_q (xP^+) w_{\bar q} ((1-x)P^+) & = & e^{-P^+/T}
\qquad {\rm (mesons)};
\\
w_q (x_1 P^+) w_{q} (x_2 P^+) w_{q} ((1-x_1-x_2) P^+) 
& = & = e^{-P^+/T} \qquad {\rm (baryons)}.
\end{eqnarray}
Only at sufficiently high $p_T$, where the parton spectrum turns into 
a power law, does fragmentation win out. Because baryons contain three 
valence quarks and their formation is suppressed in fragmentation, 
recombination remains the dominant source of baryons to higher $p_T$ 
than it is the case for mesons. 

The recombination mechanism reveals itself in a systematic difference
between the anisotropic flow patterns of mesons and baryons in 
semi-peripheral collisions. If the collective flow field exists at 
the partonic, rather than the hadronic, level the flow anisotropy for
a hadron is related to the flow anisotropy $v_2^{(q)}(p_T)$ of the 
quarks as
\begin{equation}
v_2^{(M)}(p_T) = 2 v_2^{(q)}(p_T/2) , \qquad
v_2^{(B)}(p_T) = 3 v_2^{(q)}(p_T/3) .
\end{equation}
Careful measurements for a variety of identified hadron species at
RHIC have impressively confirmed the universality of quark flow [34].

Due to the slightly increased suppression of hadrons from fragmentation
and the also increased radial flow of the expanding matter, the transition
point between recombination and fragmentation dominance is expected to
shift to larger $p_T$ (up to 10 GeV/$c$) at the LHC [35] (see Fig.~6). 
The capability of a detector like ALICE to identify hadrons well up into 
this momentum range will be a crucial prerequisite for the exploration of 
the physics of these hadrons, which are direct messengers from the 
deconfined matter. An important observable, similar as at RHIC, will be 
the species dependence of the elliptic flow, which is predicted to derive 
from a universal partonic flow pattern according to eq.~(5).

\begin{figure}[htbp]
\centering
\includegraphics[width=0.7\linewidth]{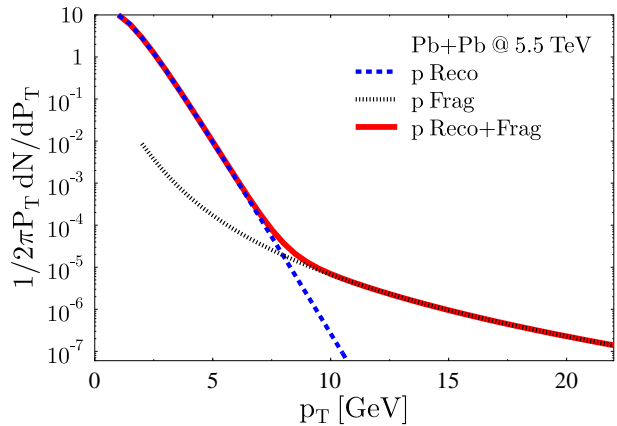}
\caption{Competition between recombination and fragmentation for proton
         emission in Pb+Pb collisions at LHC (from [35]). The calculation
         assumes a mean radial flow velocity $v_\perp/c=0.75$ for the
         thermal component.}
\end{figure}

\subsection{Heavy quarks}

Heavy quarks ($c$ and $b$) will be abundantly produced in Pb+Pb
collisions at LHC. These quarks can serve as useful probes of the
dense matter. When the color interaction in the medium is screened
at sufficiently short scales, heavy quarkonia can no longer form [36].
The required conditions for this effect have recently been reconsidered
on the basis of improved lattice calculations. It has become possible
to calculate the potential between a heavy quark pair in the color 
singlet state, which predict that the $J/\Psi$ state should survive
up to $T \approx 1.5 T_c$ in quenched QCD [37]. This conclusion is 
supported by calculations of the spectral function of a $(c{\bar c})$ 
singlet pair, which exhibits a strong peak at the location of the $J/\Psi$ 
at least up to the same temperature [38,39] (see Fig.~7). It is important 
to stress that these results are for quenched QCD, and may well change 
when dynamical light quarks are included in the lattice simulations.

\begin{figure}[htbp]
\centering
\includegraphics[width=0.5\linewidth]{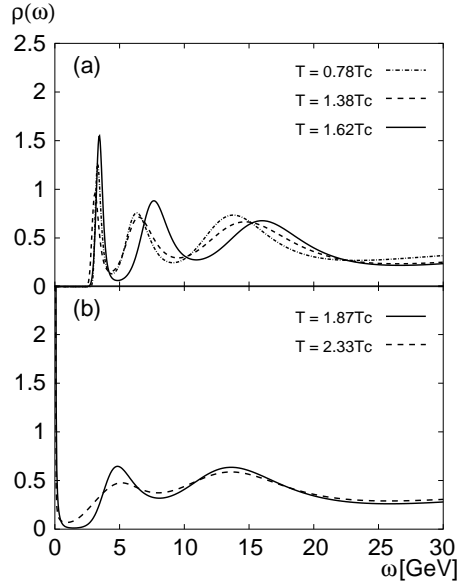}
\caption{Spectral function in the $J/\Psi$ channel for different
         temperatures below and above $T_c$ (from [38]).}
\end{figure}

On the other hand, $J/\Psi$ and $\Upsilon$ states can be destroyed 
in the hot environment by gluon bombardment, even if the states
exist [40]. Calculations predict an almost total ionization of the
$J/\Psi$ under LHC conditions and a substantial (80\%) destruction
of the $\Upsilon$ [15] (see Fig.~8). However, it is quite likely that 
($Q{\bar Q}$) bound states can be reconstituted when the plasma cools 
[41]. While the probability for this to occur is expected to still rather 
small at RHIC, it will become very substantial at LHC due to the large 
number of created $c$-quarks. Formation of $J/\Psi$ by recombination 
may even result in an effective charmonium enhancement in nuclear
collisions [15] (see Fig.~9).

\begin{figure}[htbp]
\centering
\includegraphics[height=0.49\linewidth,angle=-90]{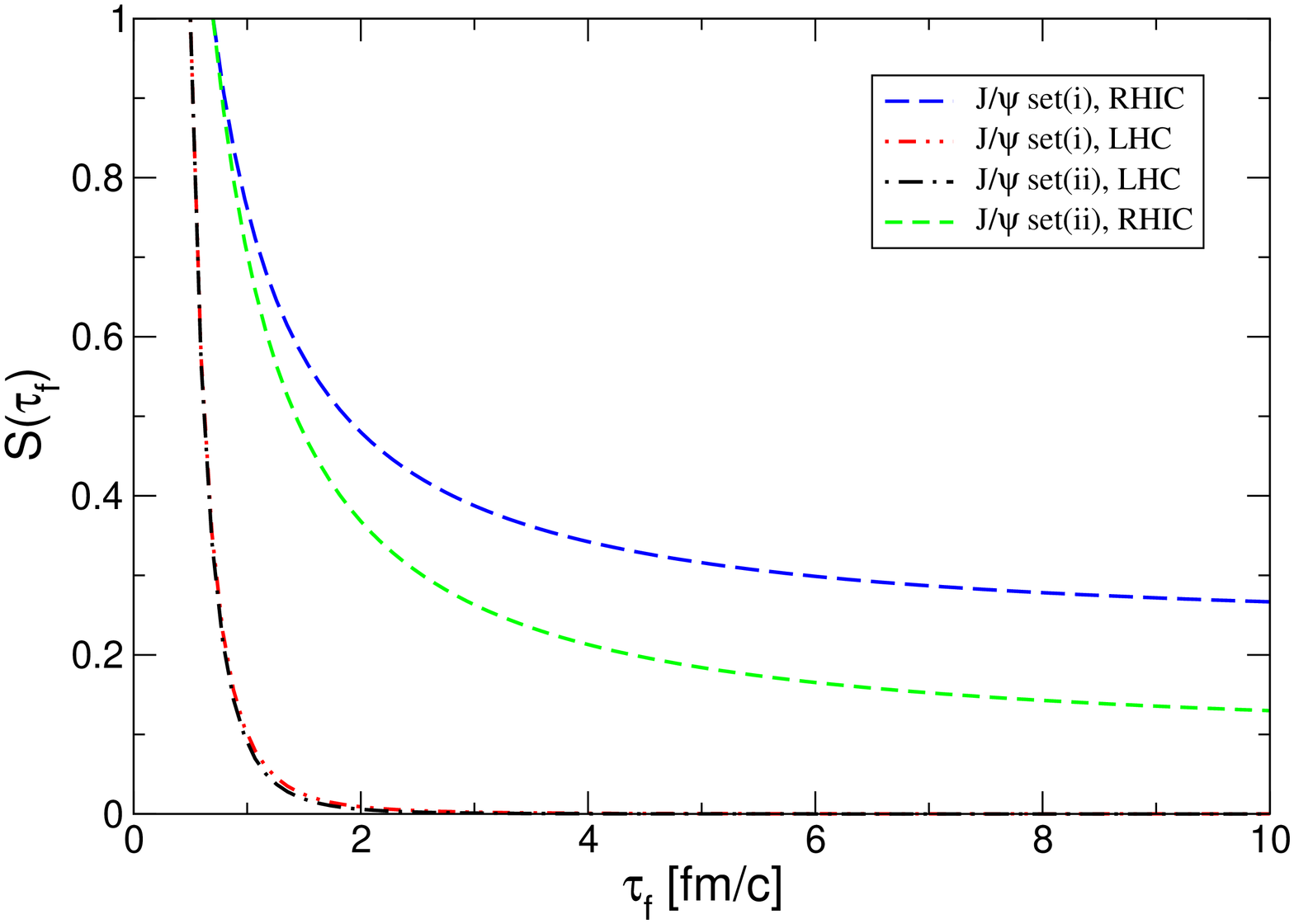}
\includegraphics[height=0.49\linewidth,angle=-90]{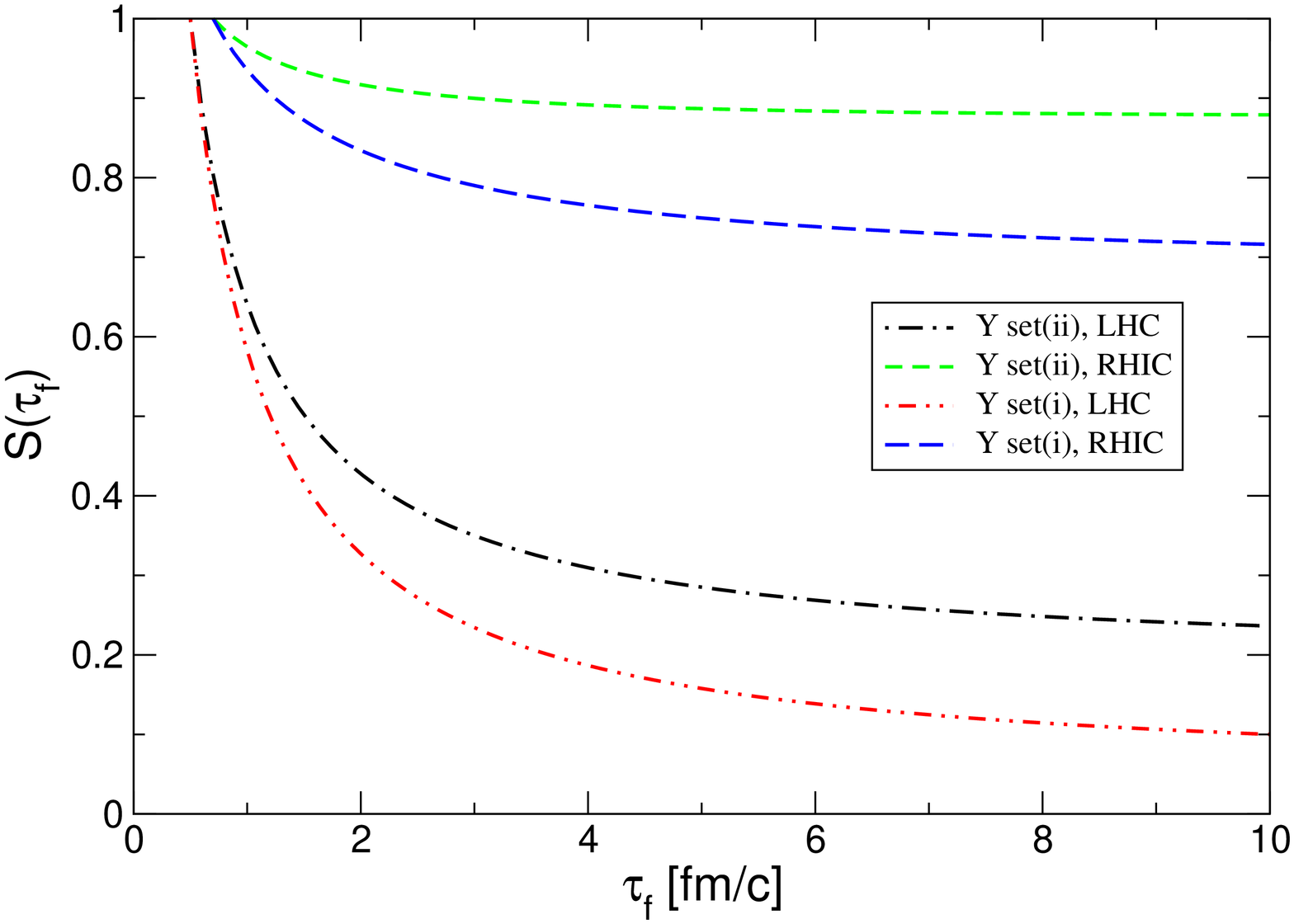}
\caption{Suppression of $J/\Psi$ and $\Upsilon$ quarkonia at RHIC and
         LHC due to inization by thermal gluons (from [15]).}
\end{figure}

\begin{figure}[htbp]
\centering
\includegraphics[width=0.65\linewidth]{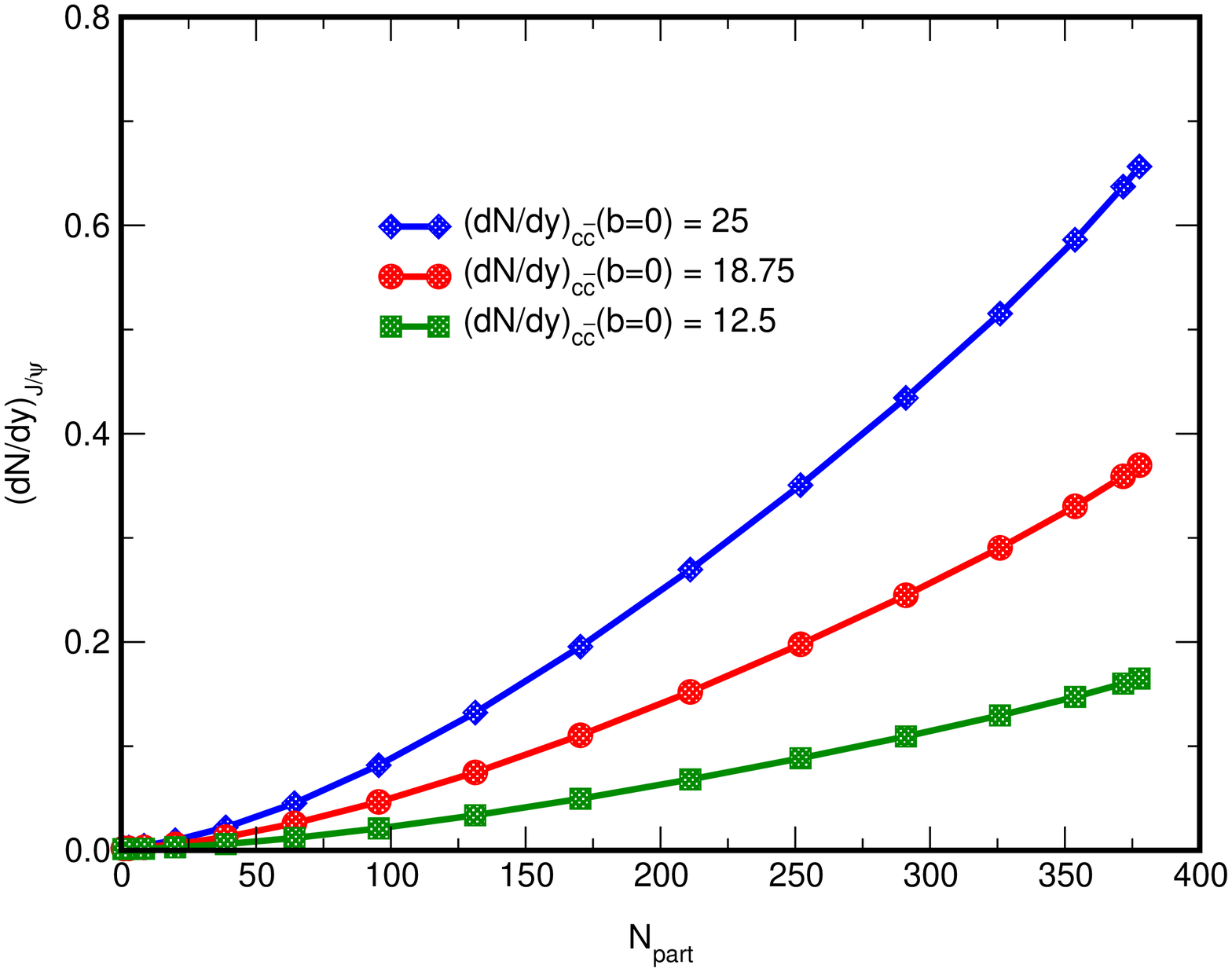}
\caption{Yield of $J/\Psi$ mesons due to recombination from a thermal,
         deconfined quark-gluon plasma for Pb+Pb collisions at LHC for
         three different assumptions of the open charm multiplicity 
         (from [15]).}
\end{figure}

\section{Summary}

I have discussed several key issues of heavy ion physics at the LHC. 
If the experiments at the SPS have brought us a first glimpse of the
quark-gluon plasma, there is every indication that enough evidence
will be collected at RHIC to announce the discovery the QGP. We are
not quite there yet, but the conditions for such a conclusion have
come into focus [13]. Assuming that this scenario will unfold
as anticipated, the LHC will become the ideal facility for a systematic 
exploration and quantitative confirmation of the insights obtained at
RHIC, aided by the plentiful abundance of hard probes. 

Some important questions, which will be at the focus of the LHC heavy 
ion program, are:
\begin{itemize}
\item How does gluon saturation work at small $x$? Is there a
      universal saturated state?
\item How does parton energy loss depend on the energy density?
      How are fragmentation functions modified in the medium?
\item Are heavy quarks thermalized in the medium? Are they deconfined?
\item What is the role of nuclear higher twist effects?
\end{itemize}
By extending the parameter range far beyond what is possible at RHIC,
the LHC will be the ideal machine to yield answers to these interesting 
and important questions.

\bigskip

{\small This work was supported in part by the U.\ S.\ Department
of Energy under grant DE-FG02-96ER40945.}

\end{document}